\documentclass[12pt,a4paper]{article}
\usepackage{graphicx}
\usepackage{amsbsy} 
\newcommand{\vs}{\vspace{-0.25cm}}

\begin{document}

\begin{center}
{\Large
\textbf{Exact calculation of three-body \\ contact interaction to second 
order}\footnote{Work supported in part by BMBF, GSI and the DFG
cluster of excellence: Origin and Structure of the Universe.}} 

\bigskip
N. Kaiser \\

\bigskip

{\small Physik Department T39, Technische Universit\"{a}t M\"{u}nchen, 
D-85747 Garching, Germany\\

\smallskip

{\it email: nkaiser@ph.tum.de}}

\end{center}

\begin{abstract}
For a system of fermions with a three-body contact interaction the second-order 
contributions to the energy per particle $\bar E(k_f)$ are calculated exactly.
The three-particle scattering amplitude in the medium is derived in closed 
analytical form from the corresponding two-loop rescattering diagram. We 
compare the (genuine) second-order three-body contribution to $\bar E(k_f)\sim 
k_f^{10}$ with the second-order term due to the density-dependent effective 
two-body interaction, and find that the latter term dominates. The results 
of the present study are of interest for nuclear many-body calculations where 
chiral three-nucleon forces are treated beyond leading order via a
density-dependent effective two-body interaction.
\end{abstract}

\bigskip


\section{Introduction and summary}
Recent advances in the formulation and construction of (low-momentum) nuclear 
interactions in chiral effective field theory have unambiguously revealed the 
important role played by three-nucleon forces \cite{evgeni,hammer,vlowkreview}. 
Three-body forces turn out to be an indispensable ingredient in accurate calculations 
of few-nucleon systems \cite{3bodycalc} as well as for the structure of light 
nuclei \cite{navratil}. In chiral perturbation theory the three-nucleon 
interaction can be constructed systematically and consistently together with the 
nucleon-nucleon potential \cite{evgeni,hammer}. At leading order it consists of a 
zero-range contact-term, a mid-range $1\pi$-exchange component and a 
long-range $2\pi$-exchange component, where the parameters of the latter component
occur also in the (subleading) $2\pi$-exchange NN-potential. The 
calculation of the subleading chiral three-nucleon force, built up from many 
pion-loop diagrams etc., has been completed recently in ref.\cite{3bodyn3lo} and 
applications  to few-nucleon systems are underway.
  
Furthermore, it has been demonstrated that by employing low-momentum two-body 
interactions (instead of traditional hard-core NN-potentials) the nuclear 
many-body problem becomes significantly more perturbative \cite{achim}. This 
desired simplification is accompanied by a prominent role of the 
three-nucleon interaction, such that its inclusion is essential in order to achieve 
(reasonable) saturation of nuclear matter already at the Hartree-Fock 
level.\footnote{At this point it should noted that three-nucleon forces are 
unavoidable, independently of whether one uses low-momentum or conventional NN-potentials
for nuclear matter calculations. Hard-core NN-potentials are simply not 
usable in perturbative many-body calculations and one has to treat them at least at the 
Brueckner-Hartree-Fock level.} The combined repulsive three-body effects counterbalance 
with increasing 
density the purely attractive contributions provided by the low-momentum two-body 
interactions alone. Improved calculations of nuclear matter which aim at 
reproducing the empirical saturation point, $\bar E_0 \simeq -16\,$MeV, $\rho_0 
\simeq 0.16\,$fm$^{-3}$, still have to treat second-order (and even higher-order) 
corrections which arise in many-body perturbation theory from the low-momentum 
two- and three-nucleon interactions \cite{hebeler}. An approximate treatment at 
second order (and beyond) is commonly pursued by mapping the three-nucleon force 
onto a density-dependent effective two-body interaction. The detailed form of the 
density-dependent effective NN-interaction as it results from the leading 
order chiral three-nucleon interaction has been worked in ref.\cite{jeremy}, in
particular with regard to its implementation into nuclear structure calculations.

The present paper aims to overcome this common approximation by performing an exact 
second-order calculation for the simplest three-body interaction, namely for the 
zero-range contact interaction. Our paper is organized as follows. In 
section 2 we recapitulate the first order calculation of the energy per particle 
$\bar E(k_f)\sim k_f^6$  and outline different methods to compute the
spin-isospin weight factors of closed three-body diagrams. In section 3 the 
two-loop diagram describing the three-particle rescattering in the medium is 
evaluated in detail. Analytical expressions are derived for the corresponding 
vacuum term $B_0$, and for the medium corrections $B_1$ and $B_2$ incorporating 
Pauli-blocking effects due to one and two particles, respectively. The ultraviolet 
divergence in the vacuum loop $B_0$ requires the introduction of a 
(Galilei-invariant) three-body counterterm proportional to fourth powers of 
momenta. The analytical results for $B_{0,1,2}$ are then used in section 4 to 
compute the second-order (i.e. five-loop) contribution to the energy per particle 
$\bar E(k_f)\sim k_f^{10}$. By comparing this novel result with the second-order 
term provided by the density-dependent effective two-body interaction, one
deduces from the numerical prefactors that the latter term actually dominates (by 
about a factor 2). The appendix contains two remarkable reduction formulas for 
special integrals over the product of three Fermi spheres.
 
The present result obtained in an exact second-order calculation with a simple
three-body contact interaction may serve to support the approximate 
treatment of chiral three-nucleon forces in nuclear matter calculations 
\cite{achim,hebeler}. Nevertheless, complete second-order calculations with the  
finite-range chiral three-nucleon forces should be attempted in the future. 
\section{First order calculation and weight factors}
\begin{figure}
\begin{center}
\includegraphics[scale=0.6,clip]{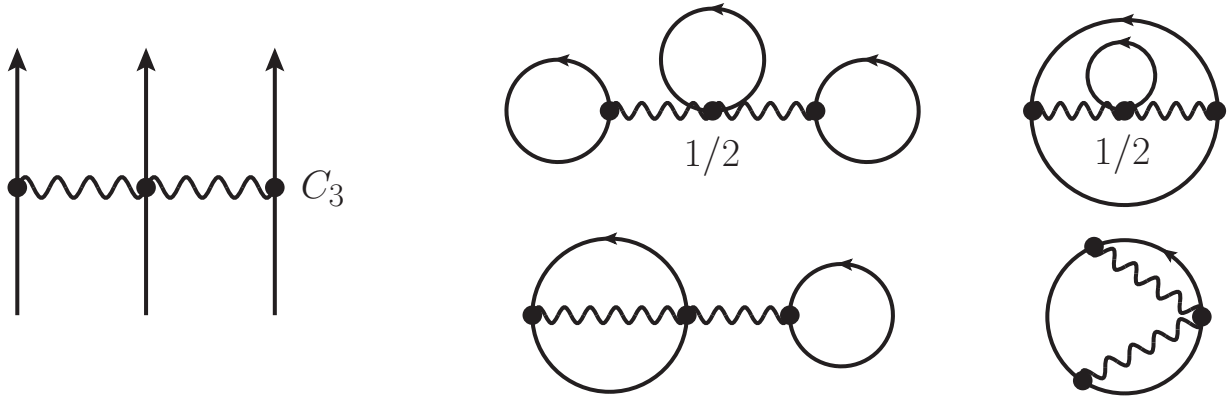}
\end{center}
\vspace{-.6cm}
\caption{Left: Three-body contact coupling modelled by heavy scalar-isoscalar 
boson exchange. Right: Closed three-loop diagrams representing the energy density 
linear in $C_3$.}
\end{figure}
We start out with reproducing the first order calculation of the energy per 
particle $\bar E(k_f)$ from a three-body contact interaction. This allows us to 
fix the notation for the coupling constant and to discuss alternative 
methods for computing the spin-isospin weight factors of closed three-body 
diagrams. The involved combinatorics becomes most transparent if one models the 
contact-vertex $C_3$ by the exchange of two heavy scalar-isoscalar bosons. The 
topologically distinct diagrams obtained by closing all three nucleon lines 
are shown in Fig.\,1 (together with symmetry factors $1/2$). The resulting 
spin-isospin weight factor in this approach is $4^3/2-4^2-4^2/2+4 =12$, and hence 
the first order contribution to the energy per particle reads:
\begin{equation} 
\bar E(k_f) = -C_3\, {3\pi^2 \over 2k_f^3} \bigg({k_f^3\over 6\pi^2}\bigg)^3
\! \cdot\! 12 =-{C_3\, k_f^6\over 12 \pi^4} = \Gamma_3\,k_f^6\,, \end{equation} 
with the density given by $\rho= 2k_f^3/3\pi^2$. The factor $k_f^3/6\pi^2$ stems 
from the volume of a Fermi sphere. For orientation and comparison we note that 
the coupling constant of the contact-term in the chiral three-nucleon force 
\cite{evgeni} is parameterized as $C_3= c_E/f_\pi^4\Lambda_\chi$. The reduced 
coupling constant $\Gamma_3=-C_3/12\pi^4$ is particularly useful 
to write down compactly the second order contributions (see eqs.(14,16)). 

\begin{figure}
\begin{center}
\includegraphics[scale=0.6,clip]{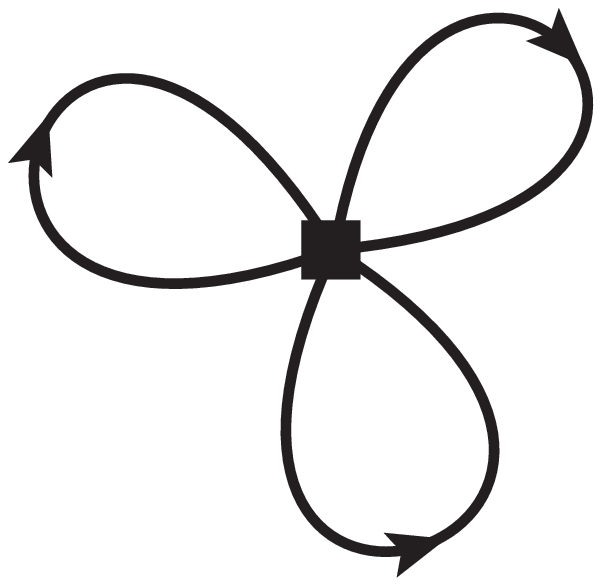}\hspace{1.0cm}
\includegraphics[scale=0.6,clip]{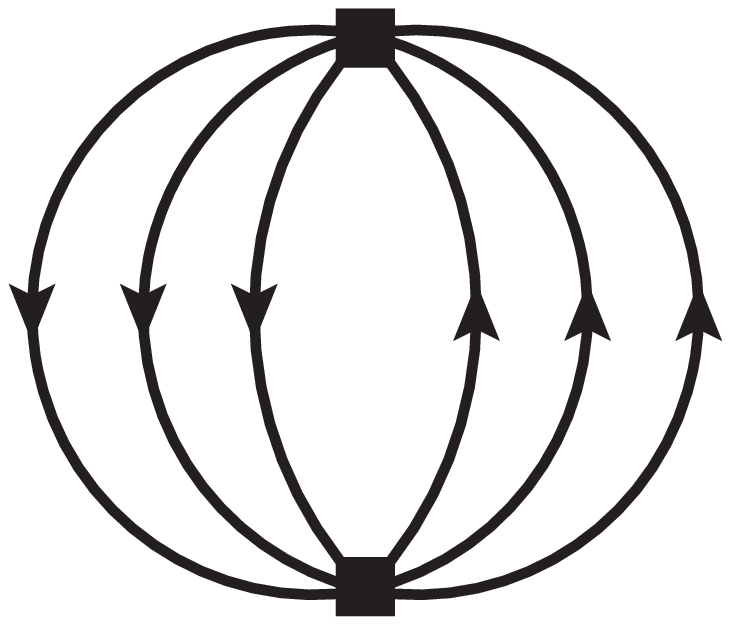}
\end{center}
\vspace{-.6cm}
\caption{First and second order diagrams generated by a three-body contact 
interaction.}
\end{figure}

An alternative derivation of the spin-isospin weight factors starts from a 
contact-vertex that additionally includes the three-particle antisymmetrization 
operator:
\begin{equation} {\cal A} = \sum_{\alpha \in S_3} {\rm sign}(\alpha) {\cal 
P}_\alpha[\boldsymbol\sigma\boldsymbol\tau]=({\boldsymbol 1}-{\cal P}_{12})
({\boldsymbol 1}-{\cal P}_{13}-{\cal P}_{23}) \,.\end{equation} 
Here, ${\cal P}_{ij} = ({\boldsymbol 1}+\vec \sigma_i \cdot\vec \sigma_j)
({\boldsymbol 1}+\vec \tau_i \cdot\vec\tau_j)/4$ is the exchange operator for 
particles $i$ and $j$, operating in their respective spin and isospin spaces. 
For the first order three-body diagram (shown in the left part of Fig.\,2) a trace 
has to be taken over the spin- and isospin degrees of freedom of all three 
nucleons, leading to the result: 
\begin{equation} 
{3C_3 \over 6}\, {\rm Tr} {\cal A} = {C_3 \over 2}\, 64  \bigg(1-{1\over 4}
-{1\over 4} -{1\over 4} + {1\over 16} + {1\over 16} \bigg)=  {C_3 \over 2}\, 64 
\cdot {3\over 8} =12  C_3\,. \end{equation}
Here, the symmetry factor $1/6$ accounts for the permutations of the three 
(indistinguishable) closed 
nucleon lines and we have displayed separately the contributions from all 
six permutations $\alpha \in S_3$. The effective coupling constant is now $3C_3$, 
since in this completely symmetric formulation of the contact-vertex each of the 
three incoming particles must be joined to the central vertex of the boson-exchange 
picture in Fig.\,1.  More interesting is the spin-isospin weight factor of the 
second order diagram shown in the right part of Fig.\,2. Taking into account its 
symmetry factor $1/6^2$, one gets: 
\begin{equation} {(3C_3)^2 \over 6^2} {\rm Tr} {\cal A}^2 =  {C_3^2 \over 4} 
{\rm Tr}\, 6{\cal A}= 36 C_3^2 =  12 C_3\cdot  3 C_3 \,,\end{equation}
where the relation ${\cal A}^2=6{\cal A}$ has been used. The factorization at the 
end of eq.(4) shows that the three-particle rescattering process carries the 
relative weight factor $3C_3$.

We finish this section by considering the density-dependent effective two-body 
interaction which results from closing one nucleon line of the three-body 
contact-vertex (see Fig.\,3). The method of employing the boson exchange picture gives 
for the two-body coupling strength: 
\begin{equation} \delta C_0(\rho) =C_3 \bigg({k_f^3\over 
6\pi^2}\bigg) \cdot 6 = {C_3 k_f^3 \over \pi^2} \,,\end{equation}
where the factor $ 6= 4+4\cdot 2 -4-2$ emerges from the set of 9 topologically 
distinct diagrams. Of course, the same result is obtained by including the 
(two-particle) antisymmetrization operator: 
\begin{equation} \delta C_0(\rho)({\boldsymbol 1}- {\cal P}_{12})\ = 3 C_3 
\bigg({k_f^3\over 6\pi^2}\bigg) {\rm tr}_3 {\cal A} = {C_3 k_f^3 \over \pi^2}
({\boldsymbol 1}- {\cal P}_{12})  \,,\end{equation}
where tr$_3$ sums over the spin and isospin degrees of freedom of the third nucleon.
Having the expression for the effective two-body coupling, $\delta C_0(\rho)=C_3 
k_f^3/\pi^2$, one can immediately give the result for its second order 
contribution to the energy per particle (see eq.(16)).
\begin{figure}
\begin{center}
\includegraphics[scale=0.55,clip]{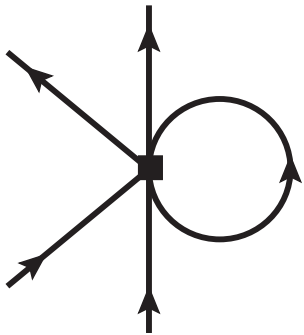}
\end{center}
\vspace{-.6cm}
\caption{Density-dependent effective 2-body interaction obtained by closing one 
nucleon line.}
\end{figure}

\section{Three-particle rescattering in the medium}
\begin{figure}
\begin{center}
\includegraphics[scale=0.55,clip]{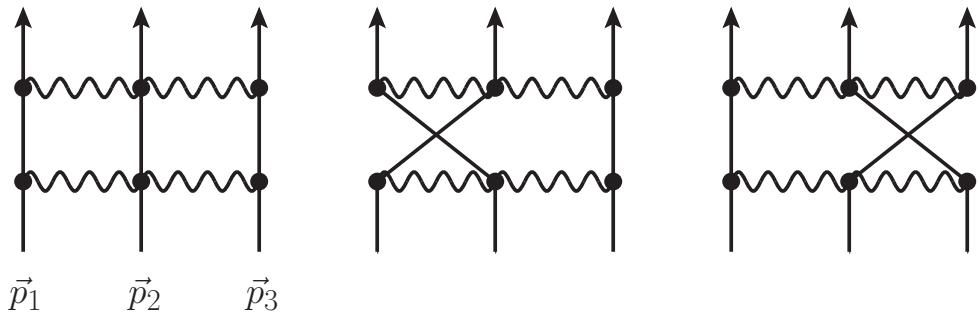}
\end{center}
\vspace{-.6cm}
\caption{Two-loop diagrams describing the three-particle rescattering in the 
medium.}
\end{figure}
The calculation of the second-order (five-loop) diagram in Fig.\,2 proceeds via the 
three-particle scattering amplitude. 
The two-loop diagrams describing the three-particle rescattering in the 
medium are shown in Fig.\,4. By including the topologically distinct diagrams with 
crossings of internal lines the crucial factor 3 deduced in eq.(4) gets accounted 
for. Each internal line introduces a particle propagator, $i[1-\theta(k_f-|\vec l_j|)]
/(l_{0j}-\vec l_j^{\,2}/2M+i \epsilon)$, and after performing the energy integrals via 
residue calculus the expression for the in-medium scattering amplitude takes the 
form:
\begin{eqnarray}  
B &=& \int\!{d^3l_1d^3l_2\over (2\pi)^6}\, {3C_3 M \over \vec l_1^{\,2}+3 \vec 
l_2^{\,2}/4 - H/6-i\epsilon} \Big[1-\theta(k_f-|\vec p+\vec l_1
-\vec l_2/2|)\Big] \nonumber \\ && \qquad\qquad \times \Big[1- 
\theta(k_f-|\vec p+\vec l_2|)\Big] \Big[1-\theta
(k_f-|\vec p-\vec l_1-\vec l_2/2|)\Big]\,. \end{eqnarray}
The chosen assignment of intermediate momenta: $\vec p+\vec l_1-\vec l_2/2$, $\,
\vec p+\vec l_2$, $\,\vec p-\vec l_1-\vec l_2/2$, with $\vec p=(\vec p_1+
\vec p_2+\vec p_3)/3$, has the advantage that interference terms of the loop 
momenta $\vec l_{1,2}$ among themselves and with the external momenta  $\vec p_j$ 
are absent in the energy denominator. The external momenta are to be taken
from the region inside the Fermi sphere  $|\vec p_j|<k_f$, hence the dimensionless 
variable $s=|\vec p\,|/k_f$ satisfies the condition $0<s<1$. The other kinematical 
quantity $H$ appearing in the energy denominator is the Galileian invariant:
\begin{equation} H = (\vec p_1-\vec p_2)^2+ (\vec p_1-\vec p_3)^2+ (\vec p_2-
\vec p_3)^2 <9 k_f^2 \,.\end{equation}
The maximum value $9k_f^2$ of $H$ is reached in the configuration where $\vec p_{1,2,3}$ 
point from the center to the vertices of an equilateral triangle of side-length 
$\sqrt{3}\,k_f$. The in-medium scattering amplitude $B$ is manifestly real-valued, since 
Pauli-blocking and energy conservation forbid any imaginary part: $3k_f^2< 
|\vec p+\vec l_2|^2+ |\vec p+\vec l_1-\vec l_2/2|^2+ |\vec p-\vec l_1-\vec l_2/2|^2 
= (2\vec l_1^{\,2}+3\vec l_2^{\,2}/2 -H/3) +\vec p_1^{\,2} +\vec p_2^{\,2}+\vec 
p_3^{\,2}= \vec p_1^{\,2} +\vec p_2^{\,2}+\vec p_3^{\,2}<3k_f^2$, where the term in 
brackets vanishes on-shell.

In the next step one expands the product of the three $(1-\theta)$-factors in 
eq.(7). This rearrangement gives a vacuum part $B_0$ with no $\theta$-factor, 
the sum $B_1$ of three equal terms with one $\theta$-factor, and the sum $B_2$ of 
three equal terms with two $\theta$-factors. The equality of the three summands is 
obvious from the structure of the two-loop diagram and can be shown explicitly by 
making the substitution $\vec l_1 = \sqrt{3}\, \vec l_3/2$. The intermediate 
momenta are then linear combinations of $\vec l_2$ and $ \vec l_3$ with 
coefficients given by rotations about angles $\pm 2\pi/3$, and the energy 
denominator $3(\vec l_2^{\,2}+  \vec l_3^{\,2})/4 - H/6$ is clearly invariant under 
these discrete rotations. At last, there is a contribution $B_3$ involving three 
$\theta$-factors. Fortunately, one does not need to evaluate the complicated term 
Re\,$B_3$, because its contribution to the energy per particle $\bar E(k_f)$ vanishes 
identically. The corresponding integral over six Fermi spheres is symmetric under 
the interchange of the external and internal momenta, except for the energy 
denominator which changes its sign. To be precise, the denominator in eq.(7)
is interpreted in this argument as a principal value.    

Next, we have to evaluate the two-loop integrals for $B_0$, $B_1$ and $B_2$.
The  divergent vacuum loop $B_0$ is treated by dimensional regularization. The 
three-dimensional integrals in eq.(7) are continued to $d$ dimensions by the rule: 
$(2\pi)^{-3}\!\int d^3 l_j \to \lambda ^{3-d}(2\pi)^{-d}\!\int d^d l_j $, which
introduces a scale $\lambda$ in order to preserve mass dimension of the loop 
integrals. The divergent behavior of the vacuum loop $B_0$ shows up through an 
Euler Gamma-function $\Gamma(1-d)$ and after expanding around $d=3$ one gets:
\begin{equation}  B_0 = {\sqrt{3} C_3M H^2\over (12\pi)^3}\bigg\{
\bigg[{1\over 3-d}-\gamma_E+\ln 4 \pi\bigg] + {3\over 2}(1+\ln 3) -\ln{-H-i 
\epsilon\over \lambda^2} \bigg\}\,.\end{equation}
A three-body counterterm proportional to $H^2$ (i.e. fourth power of momenta) is 
needed to renormalize the three-particle scattering amplitude in vacuum. This is 
different to the case of two-body scattering, where the vacuum divergence (not 
visible in dimensional regularization) can be absorbed on the scattering length 
\cite{schaefer,resum}. It is convenient to introduce via the relation $H=9k_f^2 h$ a 
second dimensionless variable $h$, which satisfies together with $s=|\vec p\,|/k_f$ 
the constraint $s^2+h<1$. Using this variable the real part of the renormalized 
vacuum loop reads: 
\begin{equation} {\rm Re}\,B_0^{(\rm ren)}= {3\sqrt{3} C_3 M k_f^4\over 
(4\pi)^3} \, h^2\bigg\{ {1\over 2}(3-\ln 3) -2\ln{k_f \over \lambda} 
- \ln h + {\rm ct}(\lambda)\bigg\}\,,\end{equation} 
with ${\rm ct}(\lambda)$ a parameter for the scale-dependent counterterm. 

In order to evaluate the two-loop integral $B_1$ we select the step-function 
$\theta(k_f-|\vec p+\vec l_2|)$. In dimensional regularization the $\vec l_1
$-integral is finite and proportional to $\sqrt{9 \vec l_2^{\,2}-2H}$. The remaining 
integral over a shifted Fermi sphere of radius $k_f$ leads to the following result 
for the real part of $B_1$:   
\begin{eqnarray} 
{\rm Re}\,B_1 &=& {3\sqrt{3} C_3M k_f^4\over (4\pi)^3}\Bigg\{ \theta
((1+s)^2-2h) \Bigg[ {\sqrt{(1+s)^2-2h}\over 10s} \Big[ 16h^2+h(9s^2-7s-16) 
\nonumber \\ && +(1+s)^3(4-s)\Big] -3h^2 \ln{ 1+s+ \sqrt{(1+s)^2-2h}\over 
\sqrt{2h}}\Bigg] \,+\, (s\to -s)\Bigg\}\,. \end{eqnarray}

For the evaluation of $B_2$ we choose the product $\theta(k_f-|\vec p+\vec l_1-
\vec l_2/2|)\,\theta(k_f-|\vec p-\vec l_1-\vec l_2/2|)$ of two  step-functions.
The  integral over $\vec l_1$-space leads to the familiar hole-hole bubble 
\cite{schaefer, resum} involving several logarithms. The condition for it 
not to vanish is $|\vec p -\vec l_2/2| <k_f$ and consequently the integration 
region in $\vec l_2$-space is a shifted Fermi sphere of radius $2k_f$. In order to 
get a more concise representation  of $B_2$ we introduce the following auxiliary 
function:
\begin{eqnarray} \boldsymbol A(Q,N) &=& 2 \sqrt{Q}\, \arctan{ N\over 
\sqrt{Q}} \,,\,\,\, {\rm for } \,\,\,\, Q>0\,,  \nonumber \\  
\boldsymbol A(Q,N) &=& \sqrt{-Q}\, \ln{ |N +\sqrt{-Q}|\over | N- 
\sqrt{-Q}|} \,, \,\,\,{\rm for } \,\,\,\, Q<0\,. \end{eqnarray} 
The final result for the real-part of $B_2$ written in terms of the variables $s$ 
and $h$ has the form:
\begin{eqnarray} 
{\rm Re}\,B_2 &=& {9 C_3M k_f^4\over (4\pi)^4}\Bigg\{ {28 \over 5}(3-s^2) 
-{22h \over 5} +\bigg[ h(7-15s^2)-3h^2-5+10 s^2 -{33 s^4 \over 5} \bigg]  
\nonumber \\ && \times  \ln\!\bigg|2s^2-h+{2\over 3}\bigg|+\bigg[ h^2 \bigg(
3+{1\over 2s} \bigg)+h \bigg(15s^2 +7s -7 -{5\over 3s}\bigg) \nonumber \\ && 
+{33 s^4 \over 5} +2s^3-10s^2- {8s \over 3} +5 +{86 \over 45 s} \bigg] 
\ln|2(1+s)^2-h|\nonumber \\ && +{16 \over 45 s} (2-3h-3s^2)^2 \Big[ 
\boldsymbol A(2-3h-3s^2,2+3s)-\boldsymbol A(2-3h-3s^2,3s)\Big] \nonumber 
\\ && +{1 \over 10 s} \Big[2h(13+s-12s^2) -21h^2 +2(1+s)^3(3s-2)\Big] 
\nonumber \\ && \times\Big[ \boldsymbol A(3(1+s)^2-6h,3(1+s))- 
\boldsymbol A(3(1+s)^2-6h,3s-1)\Big]\nonumber \\ && + {64\over 3s}
\int_0^1\!\!dx\, x \big[2(s-x)^2-h\big]\boldsymbol A\Big( 3(s-x)^2-
{3h \over 2}, 1-x\Big)    \, + \, (s\to -s)\Bigg\}\,. \end{eqnarray}
Note that the symmetrization prescription $+(s\to -s)$ applies to all terms in 
the curly brackets, also the first few which are already even functions of $s$. 
We remark again that individual components $B_j$ of the three-particle in-medium 
scattering amplitude possess an imaginary part, but these add up to zero: 
Im\,$(B_0+B_1+B_2+B_3)=0$.
\section{Results: Energy per particle at five-loop order}
\begin{figure}
\begin{center}
\includegraphics[scale=0.6,clip]{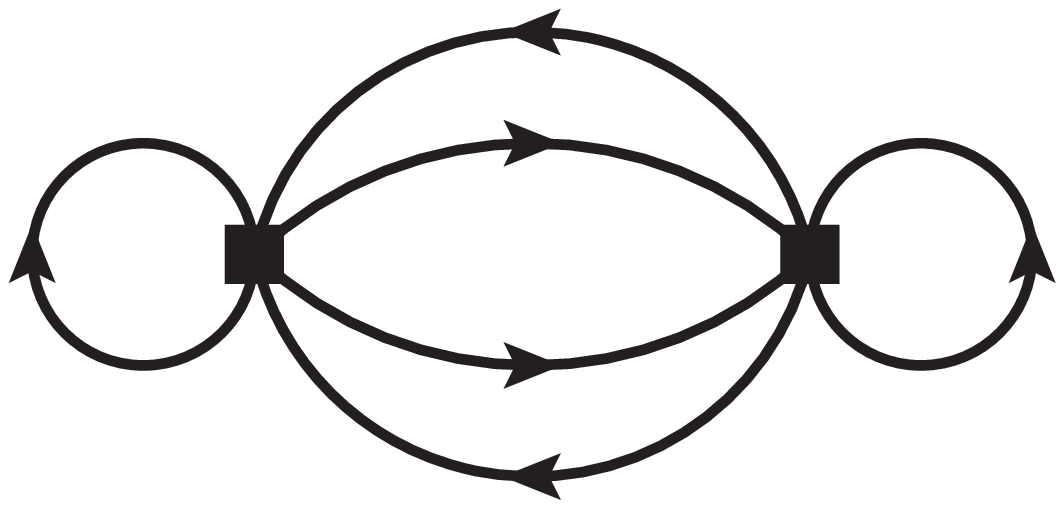}\hspace{.5cm}
\includegraphics[scale=0.42,clip]{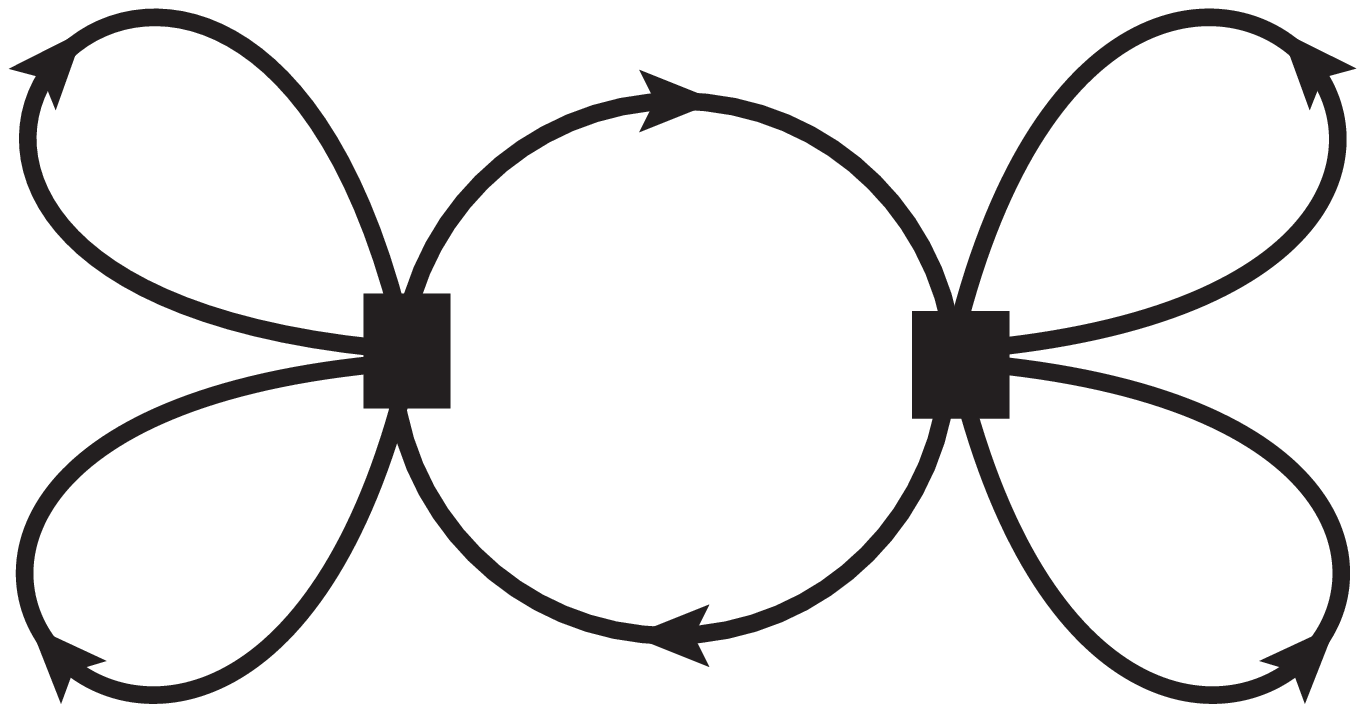}\end{center}
\vspace{-.6cm}
\caption{Left: Density-dependent effective two-body interaction to second order.
Right: Anomalous three-body diagram which vanishes at zero temperature.}
\end{figure}

Having available the analytical expressions for Re\,$B_{0,1,2}$ in eqs.(10-13) it is 
straightforward to compute the second order contribution to the energy per 
particle $\bar E(k_f)$. The pertinent integral over three Fermi spheres 
(suitably parameterized by three radii and three angles) can be solved in closed form 
for the polynomial $H^2$. Extracting the coefficient of this piece, the expression 
for the energy per particle at five-loop order (see right diagram in Fig.\,2) 
reads:   
\begin{equation} 
\bar E(k_f) =  { 37 \pi \over 175}\, \Gamma_3^{\,2} M k_f^{10} \bigg\{ \sqrt{3} 
\ln{k_f \over \lambda_0} +\zeta_0+ \zeta_1+ \zeta_2\bigg\}<0 \,,\end{equation}
where the counterterm ct$(\lambda)$ has been absorbed into the logarithm $\ln(k_f/
\lambda_0)$. The values of the numerical constants $\zeta_0$, $\zeta_1$ and $\zeta_2$ 
as obtained by integrating over three unit spheres are:
\begin{equation} \zeta_0 =-1.425\,, \qquad \zeta_1 = -5.653\,, \qquad 
\zeta_2 = -4.354\pm 0.014\,, \end{equation}
where the errors of $\zeta_0$ and $\zeta_1$ lie beyond the digits given. 
The result in eq.(14) is to be compared with the second-order contribution from 
the density-dependent effective two-body interaction as represented by the left 
diagram in Fig.\,5. Converting $\delta C_0(\rho)$ into a scattering length 
$a(\rho) = C_3Mk_f^3/4\pi^3$ and respecting the spin-isospin degeneracy factor 
$4-1=3$, one finds from the familiar low-density expansion \cite{schaefer,resum}:  
\begin{equation} 
\bar E(k_f) = { 54 \over 35}\, \Gamma_3^{\,2} M k_f^{10} \Big(11-2\ln 2\Big)>0 \,. 
\end{equation}
The numerical factors multiplying $\Gamma_3^{\,2} M k_f^{10}$ in eqs.(16,14) have 
the values $54(11-2\ln 2)/35 = 14.83$ and $37\pi(\zeta_0+ \zeta_1+ \zeta_2)/175 = 
-7.59$. Their comparison reveals that the repulsive second order contribution from 
the density-dependent effective two-body interaction is dominant, but gets reduced 
to about half of its size by the (genuine) second order three-body contribution. 
The additional logarithmic term $-1.15 \ln(\lambda_0/k_f)$ does not change this 
balance in a significant way, assuming that the scale $\lambda_0$ is of natural 
size. For the sake of completeness we note that at five-loop order one can 
additionally construct  the anomalous three-body diagram shown in the right part of 
Fig.\,5. It involves in the central loop the squared in-medium propagator and it 
vanishes (at zero temperature) by reason of the identity: $\theta(k_f -|\vec l\,|)
[1-\theta(k_f -|\vec l\,|)]=0$.

The expressions in eqs.(14,16) for the energy per particle $\bar E(k_f)\sim 
\Gamma_3^{\,2} M k_f^{10}$ at five-loop order constitute the main results of the 
present work. These analytical results demonstrate that by treating a three-body 
(contact) interaction through the density-dependent two-body interaction only
\cite{achim,hebeler} second order effects in many-body perturbation get 
considerably overestimated. Clearly, after having made this observation in a special 
case one should attempt complete second-order calculations with the finite-range 
chiral three-nucleon forces in the future.
\subsection*{Appendix: Reduction formulas for integrals over three 
Fermi spheres }
In this appendix we present two remarkable reduction formulas for integrals over 
the product of three Fermi spheres. In the first case we assume a dependence of the
integrand on the symmetric variable $s= |\vec p_1+\vec p_2+\vec p_3|/3k_f$. The 
following reduction formula holds: 
\begin{equation} \int\limits_{|\vec p_j|<k_f}\!\!\!\!\! {d^3 p_1 d^3 p_2 d^3 p_3 
\over (2\pi)^9}\, F\bigg({|\vec p_1+\vec p_2+\vec p_3|\over 3k_f}\bigg) = 
{9 k_f^9 \over 280(2\pi)^6} \int_0^1\!\! ds\, w(s) F(s)\,,  \end{equation}
with the (non-smooth) weighting function:
\begin{eqnarray} w(s) & = & 27s(1-s)^4(9 s^3+36s^2+27s-2) \nonumber \\ &&
+\theta(1-3s)\, s (1-3s)^4(54+53 s-12s^2-9s^3) \,. \end{eqnarray}
This peculiar result for $w(s)$ has been obtained with the help of Fourier 
transformation techniques and eq.(17) has been checked numerically for many 
examples of $F(s)$. 

In the second case we assume a dependence of the integrand on the combination $s^2+h = 
(\vec p_1^{\,2}+\vec p_2^{\,2}+\vec p_3^{\,2})/3k_f^2$ and derive the following reduction 
formula:
\begin{equation} \int\limits_{|\vec p_j|<k_f}\!\!\!\!\! {d^3 p_1 d^3 p_2 d^3 p_3 
\over (2\pi)^9}\, G\bigg({\vec p_1^{\,2}+\vec p_2^{\,2}+\vec p_3^{\,2}\over 3k_f^2}
\bigg) = {9 k_f^9 \over 35(2\pi)^6} \int_0^1\!\! dx\, \chi(x) G(x)\,,  
\end{equation}
with the (non-smooth) weighting function
\begin{eqnarray} \chi(x) &=& 6 \pi x^3 \sqrt{3x} + \theta(3x-1) \, 
\pi \bigg[ {1\over 4}(5-42x+105 x^2) - 18 x^3 \sqrt{3x} \bigg]\nonumber \\ && 
+ \theta(3x-2)\, \bigg\{12x^3 \sqrt{3x} \arccos {1+18 x-27x^2 \over 
(3x-1)^3} + 3 \sqrt{3x-2}\nonumber \\ && \times (5-9x +4x^2)+ (42x-5-105 x^2) 
\arctan\sqrt{3x-2} \bigg\} \,. \end{eqnarray}
This intricate expression for $\chi(x)$ has been obtained by inverting the order of 
integrations over the three involved radial coordinates $p_j^2/3k_f^2$, while fixing
their sum to $x$. Numerical checks of eq.(19) have been performed for many 
examples of $G(x)$ as well.    

\begin{figure}
\begin{center}
\includegraphics[scale=0.5,clip]{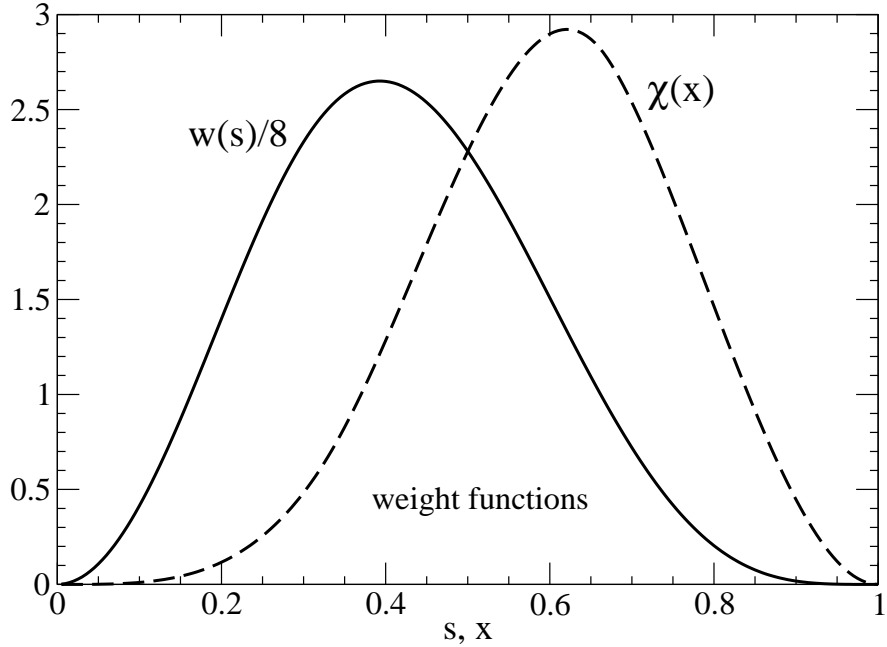}
\end{center}
\vspace{-.8cm}
\caption{Weight functions $w(s)/8$ and $\chi(x)$. The area under both curves is 
equal to $280/243$.}
\end{figure}

Fig.\,6 shows the two weight functions $w(s)/8$ and $\chi(x)$ next to each other.
The area under both (bell-shaped) curves is equal to $280/243=1.152$. A quick 
analysis gives that $w(s)/8$ reaches its maximum value of $2.65$ at $s= 0.393$, 
while $\chi(x)$ reaches its maximum value of $2.92$ at $x= 0.621$. The behavior at 
the kinematical endpoints is opposite: $w(s)=350s^2$, $s\to 0$ and $w(s)=1890
(1-s)^4$, $s\to 1$ versus $\chi(x)=6\pi \sqrt{3}\,x^{7/2}$, $x\to 0$  and $\chi(x)=
105(1-x)^2/2$, $x\to 1$.
\subsection*{Acknowledgement}
I thank J.W. Holt for informative discussions.

\end{document}